Enhancement and termination of the superconducting proximity effect due to atomic-scale defects visualized by scanning tunneling microscopy

Howon Kim[1], Shi-Zeng Lin[2], Matthias J. Graf[2], Takeo Kato[1], and Yukio Hasegawa[1*]

[1]The Institute for Solid State Physics, The University of Tokyo, 5-1-5, Kashiwa-no-ha, Kashiwa, 277-8581 Japan
[2]Theoretical Division, Los Alamos National Laboratory, Los Alamos, New Mexico 87545, USA
* e-mail address: hasegawa@issp.u-tokyo.ac.jp

abstract

Using low-temperature scanning tunneling microscopy and spectroscopy, we have studied the proximity effect at the interfaces between superconducting Pb island structures and metallic Pb-induced striped-incommensurate phase formed on a Si(111) substrate. Our real-space observation revealed that the step structures on the two-dimensional metallic layer exhibit significant roles on the propagation of the superconducting pair correlation; the proximity effect is terminated by the steps, and in the confined area by the interface and the steps the effect is enhanced. The observed results are explained quantitatively with an elastic reflection of electrons at the step edges based on calculations with the quasi-classical Green's function formulation using the Usadel equation.

When a normal metal is placed in contacted with a superconductor, superconductor-like properties appear near the interface [1-3]. Electrons in the normal metal acquire phase coherence and develop superconducting correlation due to the proximity to the superconductor. This phenomenon called the proximity effect has been utilized for designing various quantum devices [4,5] and for locally introducing superconductivity in various materials [6,7] that include topological insulators in search of the elusive Majorana fermion [8-10]. The superconducting correlation propagates into the normal metal in a scale of the coherence length. The presence of structural defects in the correlated region is thus expected to affect the propagation significantly, but it has not been known how. Here, we report on the influences of atomic-scale local structures on the proximity effect in superconductor/normal metal (SN) junctions.

At SN interfaces several transmission and reflection processes of electrons occurs. For example, when an electron in the normal metal whose energy is below the superconductor gap is injected into the superconductor, a hole is retro-reflected into the metal in order to form a Cooper pair propagating into the superconductor. This process is called the Andreev reflection. As the retro-reflected hole gains a scattering phase that depends on that of the superconductor, the electron-hole pair in the normal metal is phase-conjugated, leading to a non-zero superconducting correlation in the normal metal. As a result, superconducting properties, such as a gap at the Fermi energy, are induced in the single-particle excitation spectrum of the normal metal.

The superconducting correlation decays into the normal metal from the interface due to the scattering by impurities and defects there. Phase-breaking via inelastic and spin-flipping scattering processes limits the propagation by directly destroying the pair correlation [11]. Curiously, elastic scattering, which does not directly break the pair correlation in conventional *s*-wave superconductors, also modifies the propagation [2,3]. The scattering-induced diffusive motion confines the path of electrons near the interface and shortens their phase coherence length. Individual scattering centers, if localized, are often spatially distributed randomly and their influence on the proximity effect is renormalized into the shortened coherence length. The individual roles, thus, do not emerge explicitly, making their understanding difficult.

Using scanning tunneling microscopy and spectroscopy (STM/S), we found that individual atomic-height steps, which work as an elastic scatterer and do not break the phase correlation, have a strong influence on the induced superconducting correlation in the two-dimensional normal metal. Our real-space observations revealed that the step edges block and reflect the propagation of the superconducting correlation so that the proximity effect before (beyond) the step edges is enhanced (terminated). The observed results are explained quantitatively with an elastic reflection of electrons at the step edges based on theoretical calculations using the Usadel equation. The results demonstrate a linkage of the atomic structure with the microscopic functionality, and open up the possibility of controlling the superconducting correlation and eventually fabricating new devices with novel properties.

The STM/S is an ideal tool to investigate the spatial distribution of the proximity effect, since it is capable of measuring the local density of states (LDOS) with spatial resolution less than 0.1 nm [12], and the methods have been already utilized in various SN systems [13-17]. Here we investigate it between crystalline lead (Pb) island structures and a Pb striped-incommensurate (SIC) phase using a $^3$He-cooled low-temperature STM system. A metallic SIC phase was prepared by 1.3 ML-Pb deposition onto the Si(111)7×7 surface at room temperature, followed by annealing at 640~660 K for 3 min. [18]. In order to form superconducting Pb islands, Pb was deposited on the SIC phase at 240 K. Before transferring the sample to the low-temperature STM unit, the sample was kept at room temperature to make the top of the islands flat.

The SIC phase becomes superconducting below the critical temperature, which is 1.83 K and 1.1 K according to STM [19] and electrical conductance measurements [20], respectively. We performed the experiments at 2.15 K so that the SIC phase is normal, as is confirmed by the tunneling spectra as shown in Fig. S1(b) of the Supplementary Materials (SM). In order to investigate the spatial distribution of the gap, we map out the tunneling conductance at the zero-bias voltage, which we call the zero-bias conductance (ZBC) [21,22], corresponding to the LDOS at the Fermi energy. For quantitative analysis of the spatial variation, we plot the negative ZBC normalized by that measured on the normal metal far from the Pb islands. The normalized negative ZBC value is 0 (-1) for a perfect (no) gap, corresponding to the superconducting (normal metal) area, respectively.

Spatial variation of tunneling spectra taken around the edge of Pb islands (See Fig. S2 in the SM.) clearly demonstrates the leaking of the superconducting correlation with gap-like features at the Fermi energy in the normal SIC phase. The gap depth decays from the SN interface with the decay length of 40.5 ± 1.7 nm in our measurements. The length is slightly longer than that reported by Kim *et al*. [16], which can be explained by the lower temperature of our measurements. The decay length, obviously shorter than that expected for the ballistic regime (>500 nm), and thus limited by the elastic scattering, indicates the diffusive nature of the metal layer. On the superconducting side, on the other hand, no deterioration in the superconducting gap is observed even close to the interface, which is explained by a large conductance difference between the crystalline Pb island and the atomically-thin metal layer. [16,17]

Since our normal metal is a two-dimensional layer formed over the semiconductor substrate, steps on the substrate have a significant influence on the diffusion of electrons. In order to elucidate the influence of the steps on the proximity effect, we measured the tunneling conductance around Pb islands formed in a stepped area of the silicon substrate. Figure 1(a) is an STM image showing 9~22 monolayer (ML) -high Pb islands, elongated along the step-edge direction. The steps go down from the left to the right side of the image. Since the growth of the islands tends to be terminated at the substrate steps, the islands's edges often overlap with the substrate step edges. A ZBC map taken in this area is presented in Fig. 1(b). All Pb islands are colored green, which indicates a vanishing ZBC and good superconductivity there, and the normal metal far from the Pb islands is colored yellow, indicating no gap there. The area surrounding the Pb islands has the color of blue to red, implying that a weak gap is induced by the proximity effect.

One thing that we notice from the ZBC map is that the steps, which are marked by dashed lines, terminate the propagation of the proximity effect into adjacent normal metal. The cross-sectional plot along the line aa' in Fig. 1(c) demonstrates the decay of the ZBC in the normal metal region and its sudden disappearance beyond the step edge. Recent critical current measurements on atomically-thin superconducting layer by Uchihashi *et al.*

[23] indicate that the steps serve as a blocking barrier. We speculate that the barrier also limit the propagation of the pair correlation and thus terminating the proximity effect there, which will be discussed later.

One may also notice in the ZBC map (Fig. 1(b)) that the colormap in the normal metal layer, just outside of the edge of Pb islands, is not uniform and has significant variation, which is in contrast with the cases of previous studies on flat metallic layers [16,17]. The sites marked A in the ZBC map, where the edges of the Pb island are situated directly on the flat SIC phase, have the normalized ZBC of -0.33 (See Fig. S3 in the SM for statistical details), as shown in the ZBC plot of Fig. 2(a). The site B (C), where the edge of the island and the upward (downward) step of the substrate coincide, has the normalized ZBC of -0.61 (-0.80). The difference in the ZBC across the interface is a measure of the transparency of the interface, that is, electrical conductivity through the SN interface. [24] The SN interfaces with a large energy barrier suppress the Andreev reflection, while they enhance the elastic scattering of electrons at the interface, and thus the superconducting correlation into the SIC phase is suppressed. The observed ZBC difference among the sites can be explained with different interfacial conductivities that depend on the local and atomic structures of the interface.

As schematically shown in Fig. 2(b), because of the direct contact, the energy barrier of the site A for electrons moving from the Pb island to the SIC phase should be small. At site B, which is often observed on the upper side (left side in the case of Fig. 1) of the islands, an atomically thin contact is formed between them. At site C, on the other hand, the Pb island is separated from the SIC phase by the atomic step. In order to explain the observed interface dependence quantitatively, we performed a numerical analysis based on the quasi-classical Green's function formulation using the Usadel equation [25-27], which has been utilized for the proximity effect in diffusive metals (See S5 in SM for details.), including the present system [16]. The fitting results, by just changing the interface resistance for the three cases, shown in Fig. 2, support the above-mentioned explanation.

In the upper-left of the STM image in Fig. 1(a), there is a Pb island that is directly contacted with the SIC phase underneath, same as the case of site A. As a downward step edge is close to the island, the terrace width of the sandwiched SIC phase is quite narrow, less than the decaying length in the metal layer (~ 40 nm). We found that in such a confined area the proximity effect is enhanced, as shown in the ZBC map of Fig. 3(a). Figure 3(b) shows several cross-sectional plots taken in the areas with various terrace widths. The plots clearly show the gap depth in the confined area is larger than that measured on the flat terrace, and it is larger (smaller ZBC) for a narrower terrace. For instance, the ZBC at the normal metal side of the interface is ~ 0.2 for the narrowest terrace (12.8 nm width), obviously having a larger gap than the wide-terraced area (*eg*. site A, whose ZBC is ~0.4). The inset of Figure 3(b) shows a plot of the ZBC measured at the normal metal side of the interface along the periphery of the Pb island (yellow line in Fig. 3(a)). The plot shows a gradual decrease of the ZBC with the terrace width, providing further evidence for the enhanced proximity effect in the geometrically confined region.

The enhanced proximity effect is explained by the elastic scattering of electrons and holes at the step edge, since the elastic scattering does not break the pair correlation and the reflected pair enhances it. Multiple Andreev reflection also contributes to the enhancement; electrons and holes redirected by the scattering are again injected to the SN interface to generate the pair correlation. The modification of DOS by the confinement should be negligible because the normal metal is diffusive. In fact, we did not observe any modulated LDOS in the confined area in the ZBC map taken under the perpendicular magnetic field of 0.17 T to suppress both superconductivity in the Pb islands and the proximity effect (See Fig. S4 in the SM).

In order to obtain theoretical support for the above-mentioned scenario, we calculated the ZBC profiles for various terrace widths using the Usadel equation with appropriate boundary conditions. The calculated structure has three regions; a superconductor / normal metal / normal metal, (SNN) as depicted in Fig. 3(c). The central normal metal has a length of the terrace width of the confined area. In this model, we use two interfacial conductivities to characterize the transparency of the two interfaces. Here $g_{SN}$ is the interfacial conductivity at the SN interface and was set to a value for site A, whereas $g_{NN}$ is the interfacial conductivity of the abutting normal metals. All other parameters, which are related to the conductivity and the diffusion constant of each region, were set to the same value as Fig. 2(a). As shown in Fig. 3(b), the experimental results can be fitted well; the normalized negative ZBC is larger in the confined area, compared to the flat area, and becomes larger in the narrow terrace. The fitted $g_{NN}$ is quite small compared with $g_{SN}$, indicating that significant reflection of electrons at the NN step edge indeed contributes to the enhanced proximity effect. The small $g_{NN}$ also explains the termination of the proximity effect across the NN step edges, as demonstrated in Fig. 1(c). It is found that the estimated interfacial conductivity $g_{NN}$ are consistent with that estimated by Uchihashi *et al*. [23] (For more details, see S5 in the SM).

In conclusions, using low-temperature STM/S, we have studied the proximity effect at the interfaces between superconducting Pb islands and two-dimensional metallic SIC layer. We found that the step structures on the SIC phase can terminate and enhance the propagation of the superconducting pair correlation depending on their configuration. The observed results will provide a guide for the effective introduction of superconducting properties into various non-superconducting materials, which aims for unique physical states in the mesoscopic scale.

We appreciate fruitful discussion with Masaru Okada, Kazuo Ueda, Nobuhiko Hayashi, Mahn-Soo Choi, Dimitri Roditchev, Michael Tringides, and Ing-Shouh Hwang. This work was supported by JSPS KAKENHI Grant Number 25286055. Work at Los Alamos National Laboratory was supported through the U.S. DOE contract No. DE-AC52-06NA25396 by the LDRD program (S.Z.L.) and by the Office of Science, Division of Materials Sciences and Engineering (M.J.G.).

**Figure captions**

**Figure 1.** (a) STM image of Pb islands formed on a SIC-phase covered surface (1.0 μm square, $I_T$ = 50 pA and $V_S$ = 50 mV). The edges of the Pb islands and the steps of the SIC phase are highlighted with white and black dotted lines, respectively. (b) The zero-bias conductance (ZBC) color map of the same area as in (a). (c) Normalized negative ZBC (upper) and topographic profiles (bottom) taken along the line aa' drawn in (b). The vertical dotted lines colored red, green, and blue corresponds to the positions of the normal metal step, the SN interface, and a step on the superconductor, respectively.

**Figure 2.** (a) Normalized negative ZBC profiles taken at three different sites, A, B, and C. Thick solid lines indicate fitted curves by the Usadel equation. The fitting parameter, $\tilde{g}_{int}$, which is proportional to the electrical conductivity through the interface, is 0.80, 0.30, 0.11 for sites A, B, and C, respectively, whereas the other parameters are fixed ($D_{SIC}/D_{Pb}$ = 0.04, $\sigma_{SIC}/\sigma_{Pb}$ = 0.21). (b) Schematics of the three different sites.

**Figure 3.** (a) 400 nm × 400 nm ZBC color map taken on a confined area surrounded by the Pb island and a step edge of the SIC phase. The edges of the Pb islands and the SIC steps are highlighted with white and black dotted lines, respectively. (b) Normalized negative ZBC profiles across the SN interface and the step edges measured along the colored lines drawn in (a). The length written for each plot is the terrace width measured along the corresponding line. The light-colored thick lines are fitted curves based on the Usadel equation calculated in the SNN model. The fitting parameters that correspond to the conductivity through the SN interface, $\tilde{g}_{int\,SN}$, and

through the NN interface, $\tilde{g}_{\text{int}\,NN}$, are 0.64/0.20, 0.80/0.26, 0.56/0.19, 0.56/0.11, 0.55/0.10 in a sequence of the terrace width from 12.8 nm to 76.7 nm. (Inset) normalized ZBC profile along the yellow line in (a), which is drawn parallel and 5 nm away from the edge of the Pb island. Red line is a guide to the eye. (c) Schematics of the SNN model.

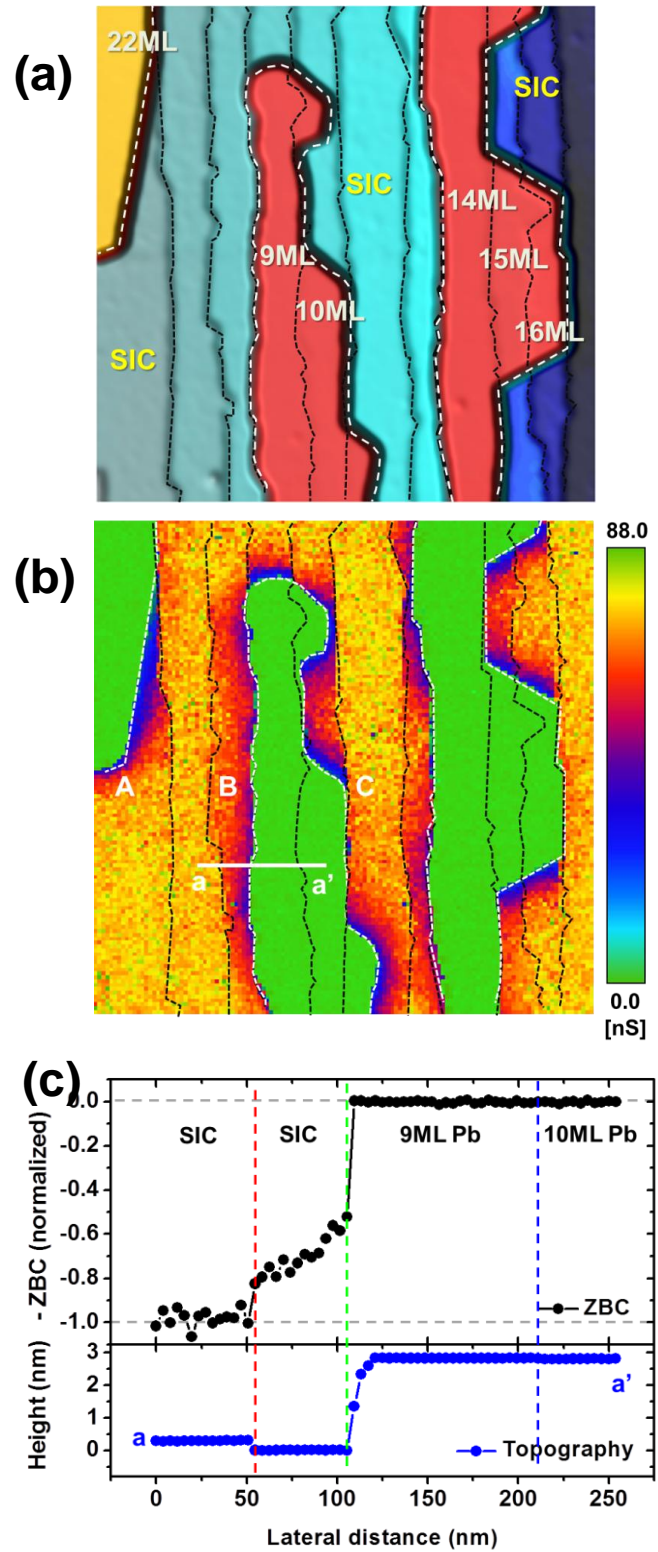

Figure 1

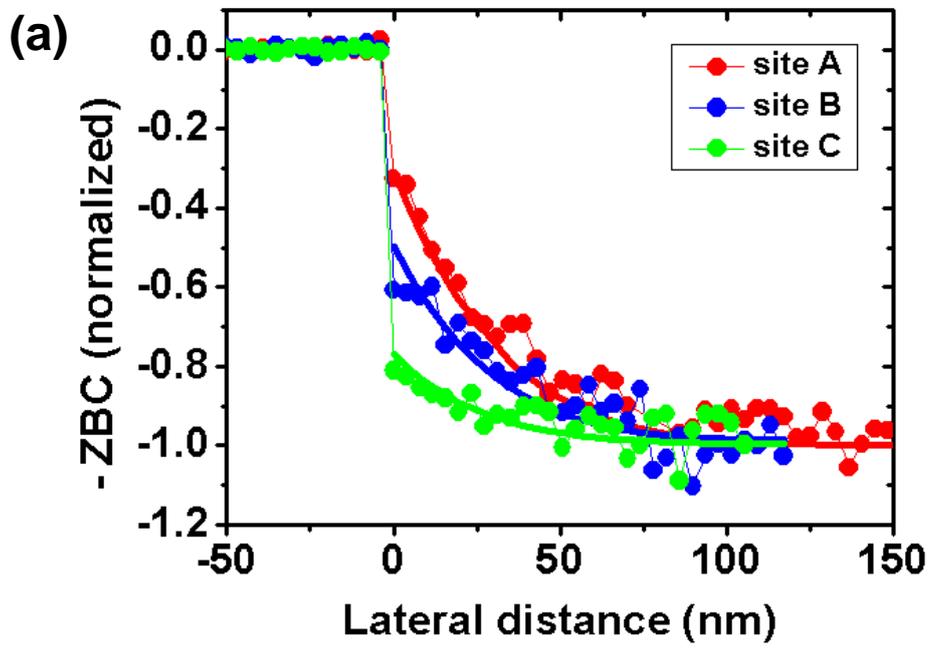

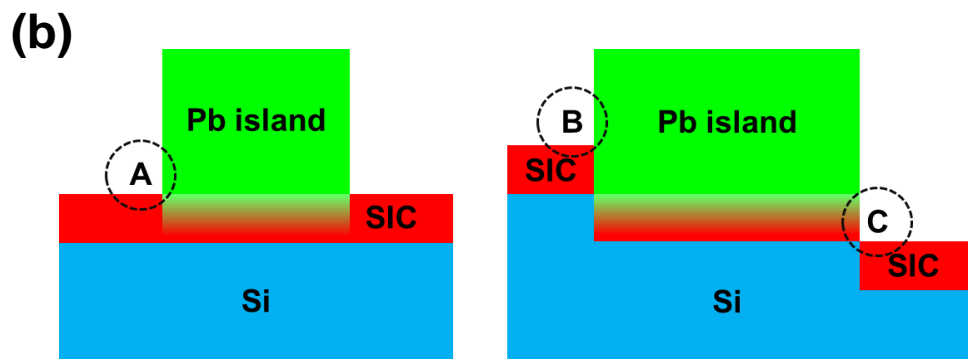

Figure 2

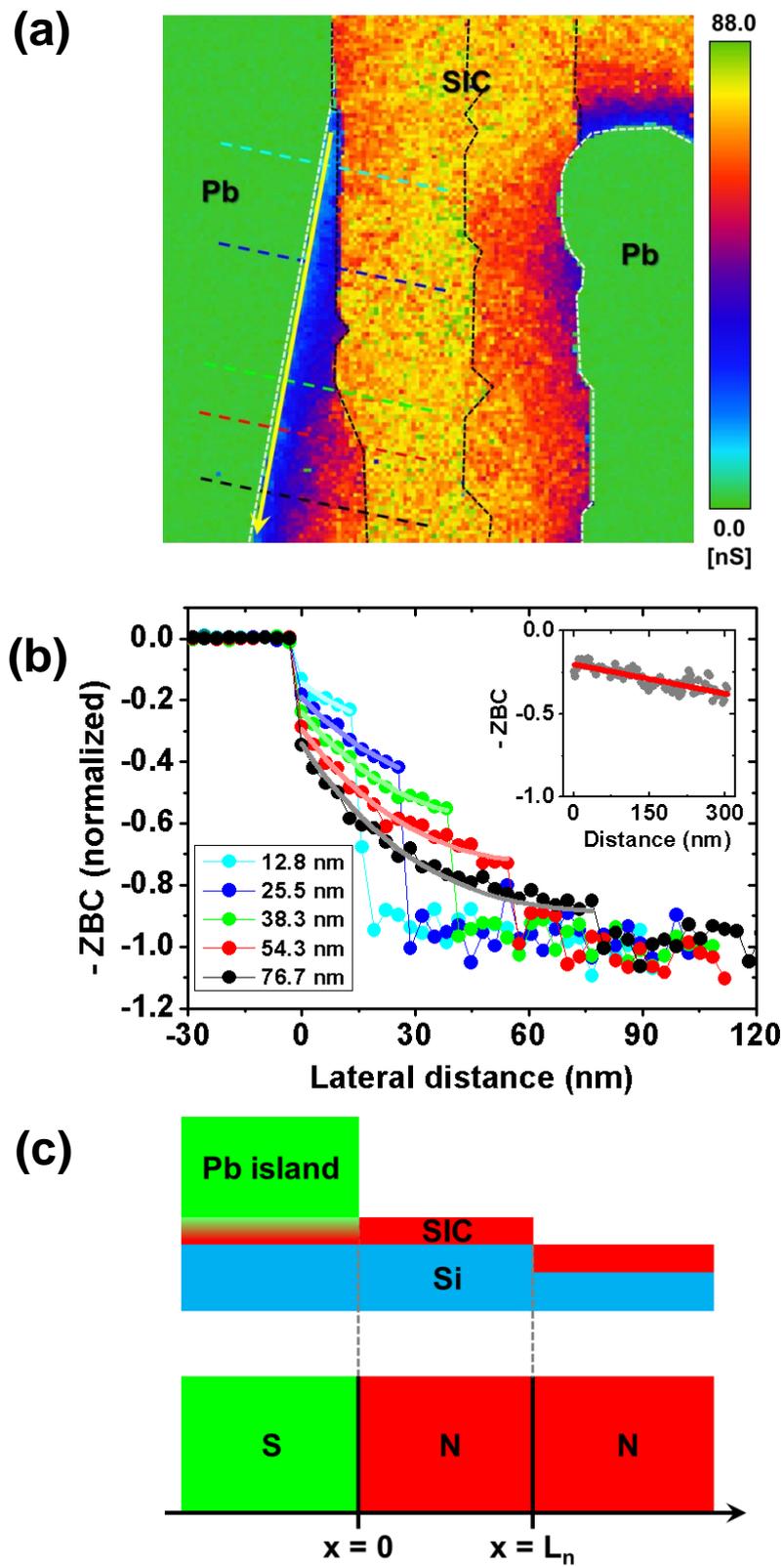

Figure 3

Supplementary materials of
"Enhancement and termination of the superconducting proximity effect due to atomic-scale defects visualized by scanning tunneling microscopy "
by Howon Kim, Shi-Zeng Lin, Matthias J. Graf, Takeo Kato, and Yukio Hasegawa

## S1. Structure and electronic properties of Pb islands and the SIC phase

The samples were prepared *in situ* in situ sample preparation chamber of a 3He-cooled low-temperature STM system (Unisoku, USM-1300S with a Nanonis controller). As a substrate we used (111)-oriented Si wafer (As doped, 1-3 mΩ•cm). A metallic SIC phase was prepared by deposition of 1.3 ML of Pb (99.9999%) onto the Si(111)7×7 surface at room temperature, followed by annealing at 640~660 K for 3 min [1]. In order to form superconducting Pb islands, Pb was deposited on the SIC phase at 240 K with a deposition rate of 1.5 ML/min. Before transferring the sample to the low-temperature STM unit, the sample was kept at room temperature for 90 min to make the top of the islands flat.

An STM image taken on a sample prepared mentioned above is shown in Fig. S1. Mechanically polished PtIr tips (Unisoku) were used for all STM/STS measurements in this study. A part of a Pb island is observed on a terrace that is covered with the Pb-induced striped-incommensurate (SIC) phase. In the SIC phase, domains that have a √3×√3 structure are separated by domain walls, which have a √7×√3 structure locally and contrasted dark in the image [2,3].

The thickness of the Pb island is 14 monolayer (ML) counted from the SIC terrace (1 ML = 0.284 nm for Pb(111)). According to x-ray diffraction study [4], the SIC structure under Pb islands is converted to bulk Pb, and therefore, for the real Pb thickness 1 ML should be added to the thickness counted from the SIC phase. A Moiré pattern is observed on the Pb island, which is caused by strain due to the lattice mismatch between Pb and Si at their interface [5].

Superconductivity of the Pb island and the SIC phase was characterized by tunneling spectra. All differential tunneling conductances in this study were measured in a standard lock-in method whose modulation is 100 μV$_{rms}$ at 2 kHz, while the tip-sample tunneling junction was stabilized with the sample bias voltage of 5 mV and tunneling current of 400 pA. The spectra taken at 0.5 and 2.1 K are shown in Fig. S1(b). The superconducting critical temperature of Pb thin film changes as $T_c(1-d_c/d)$, where $T_c$ is the critical temperature of bulk Pb (7.2 K) and $d_c$ is the critical thickness (1.57 ML) [6]. The critical temperature estimated from the formula for a 14-ML Pb thin film is 6.4 K. Since it higher than the measurement temperatures, the spectra taken at both temperatures show the superconducting gap, and both can be fitted well with the Dynes function [7], which is a modified Bardeen-Cooper-Schrieffer (BCS) function with the thermal broadening function $\Gamma$.

$$\frac{dI_{ns}}{dV} = \rho_t(0)\rho_n(0) \int_{-\infty}^{\infty} \text{Re}\left\{\frac{|E - i\Gamma|}{\sqrt{((E - i\Gamma)^2 - \Delta^2)}}\right\} \left[\frac{\exp[(E + eV)/k_B T]}{k_B T \{1 + \exp[(E + eV)/k_B T]\}^2}\right] dE$$

From the fitting curves shown with a red line in Fig. 1(b), we found the gap $\Delta$ of the 14-ML Pb island is 1.24 meV with $\Gamma = 0.003$ meV at 0.5 K, and 1.08 meV with $\Gamma = 0.095$ meV at 2.1 K. In the present experiments, we studied the proximity effect around Pb islands whose thickness ranges from 9 to 22 ML, and all islands have the critical temperature higher than 5 K.

The critical temperature of the SIC phase is 1.83 K [8] / 1.1 K [9]. We obtained the tunneling spectra showing the superconducting gap below the critical temperature. The gap obtained from the fitting is 0.29 meV with $\Gamma = 0.01$ meV at 0.5 K.

These results indicate that the Pb islands are superconducting and the SIC phase is a normal metal at the measurement temperature (2.1 K).

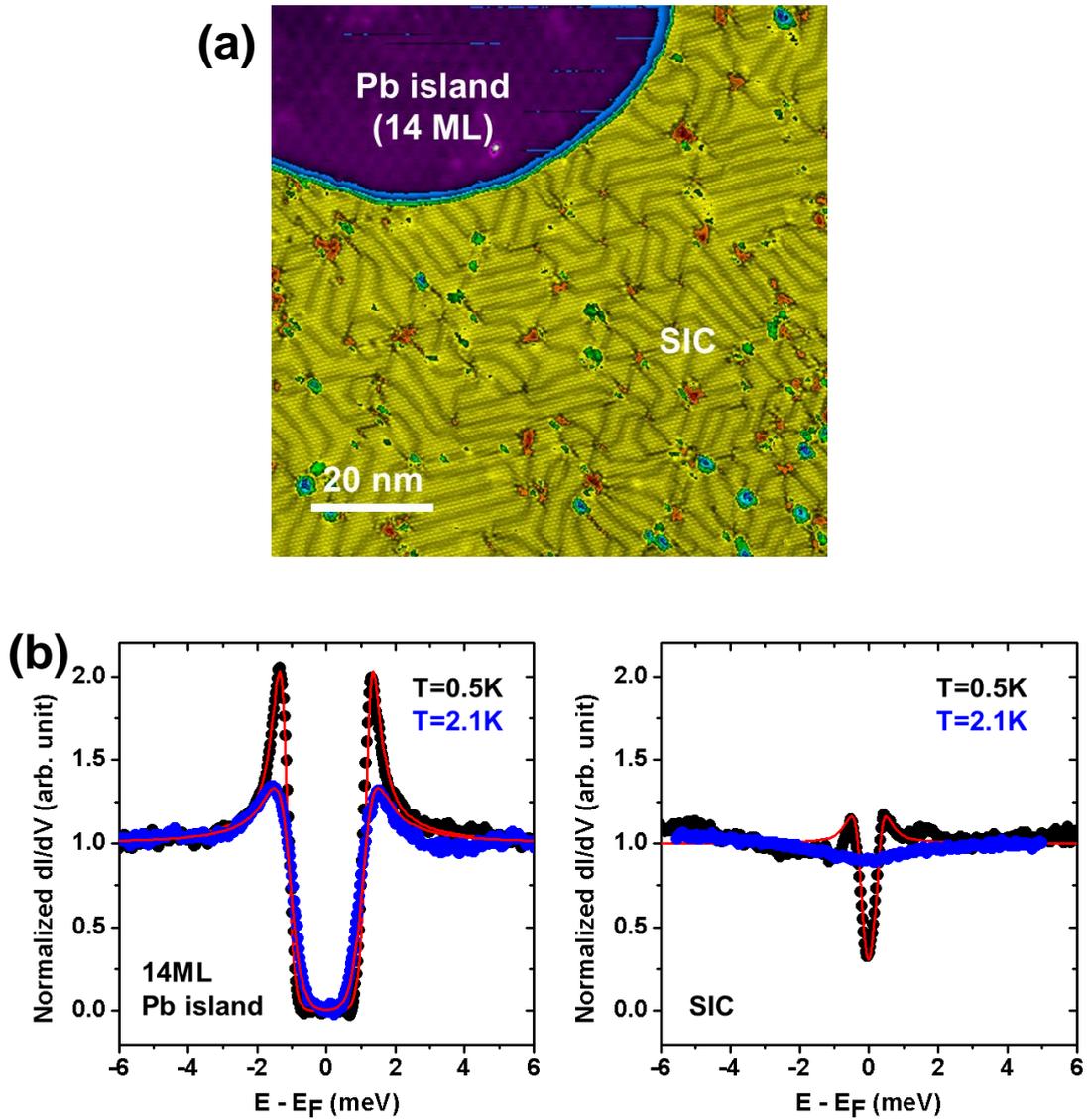

**Figure S1.** (a) STM image of a Pb island (14 ML thickness) grown on SIC phase/Si(111) surface. The image size is 90 nm × 90 nm and the tunneling conditions are $I_T$ = 50 pA and $V_S$ = +2 V. (b) Differential conductance taken on the 14 ML-thick Pb island (left) and the SIC phase (right) below (black curves) and above (blue curves) the critical temperature of the SIC phase ($T_C$=1.8 K). All spectra are normalized by the conductance outside the gap at $V_S$ = 6 mV. The red lines are fitting curves with the Dynes function.

## S2. Proximity effect around an interface between a Pb island and flat SIC phase

We investigated the proximity effect between a Pb island and the SIC phase by taking the tunneling spectra across the interface between them. Figure S2(a) is an STM image showing an 8 ML Pb island on the SIC phase. Color-coded tunneling spectra (Fig. S2(b)) taken along the line shown in the STM image indicate superconducting gap on the Pb island and no gap on the SIC phase far from the island. On the SIC phase near the island, however, a weak gap due to the proximity effect is observed, as more clearly demonstrated in Fig. S2(c) by the spectra taken at the 3 sites; on Pb island (A), SIC phase near the island (B), and SIC phase far from the island (C).

In the tunneling spectra taken on the normal metal far from the island; spectrum C in Fig. S2(c), we found a small dip at the zero bias voltage. A similar dip was reported in the tunneling spectra above the upper critical field ($> 0.145$ mT) of the superconducting SIC phase [8]. The dip is presumably due to dynamical Coulomb blockade as recently demonstrated by Brun *et al.* [10,11] In order to eliminate the dip effect on the observation of the proximity effect, we normalized measured ZBCs with a ZBC which was obtained 200 nm away from the island. A plot of the normalized negative ZBC in Fig. S2(d) shows the decay of the proximity effect away from the island, and from an exponential fit, we found the decaying constant is $40.5 \pm 1.7$ nm.

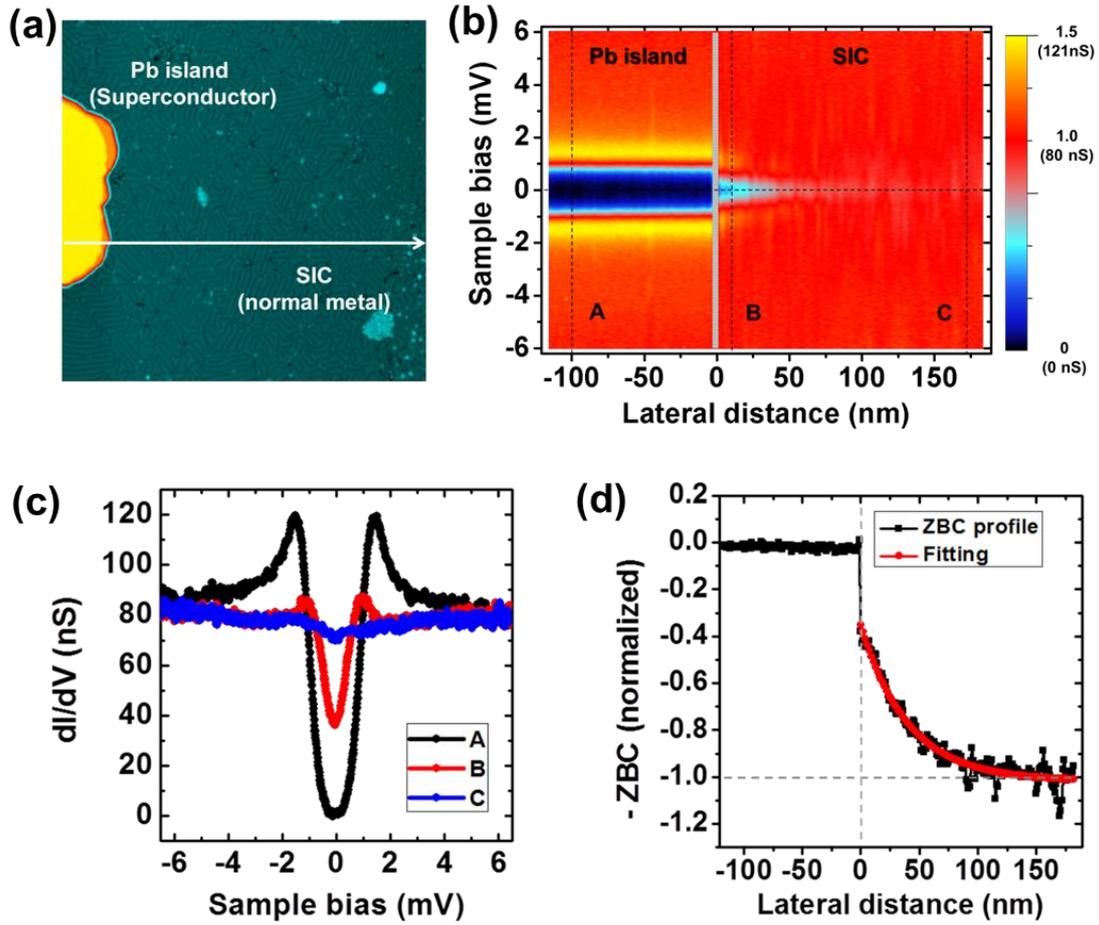

**Figure S2.** (a) 200 nm × 200 nm STM image showing an interface between a 8-ML Pb island and SIC phase on Si(111). The imaging condition is $I_T$ = 30 pA and $V_S$ = +50 mV. (b) color-coded 300 tunneling spectra taken along the 300 nm line from the Pb island to the SIC phase shown in (a). All spectra are normalized by zero bias conductance taken at the SIC surface far from the interface (80 nS). (c) Tunneling spectra extracted from A (on the Pb island), B (on SIC near the interface), and C (on SIC far from the interface). (d) ZBC profile. Red line is an exponential fitting curve of $y$ = -1.02 + 0.67exp(–$x$/40.5 nm).

## S3. Distribution of normalized ZBC measured at the normal metal side of the interface

Figure 3 is the normalized negative ZBC measured on the SIC phase in the vicinity of the Pb islands. As mentioned in the main text and described in Fig. 2(a), there are three types of interface between the Pb islands and the SIC phase, and because of their different transmission probability, their ZBC values at the SIC phase side of the interface are also different. In this analysis, we did not include the points at which the terrace width of SIC is narrow ($< 150$ nm).

From the standard deviation of the histogram, we obtained a margin of error for each site, as written in the main text.

Please note that the maximum ZBC value of site A does not exceed 0.2 when the SIC terrace is wide, sharp contrast with the case of the narrow terrace as shown in Fig. 3.

We did not find significant dependence of the ZBC on the thickness of the Pb islands.

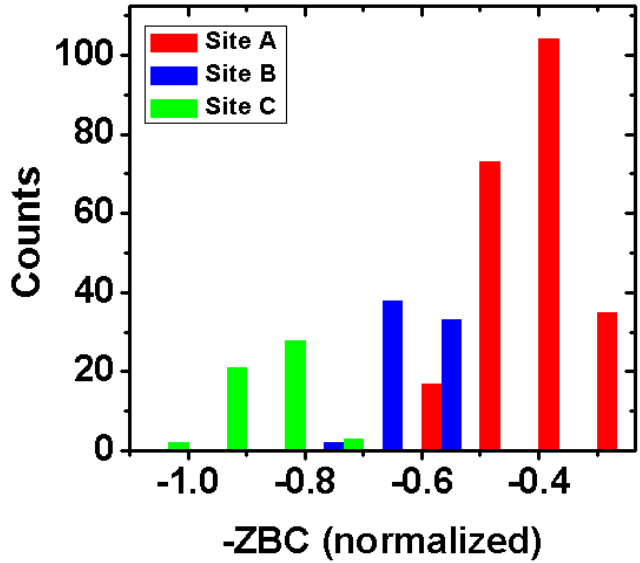

**Figure S3.** A statistical histogram of the normalized negative ZBC values on the SIC phase that are taken at the closest distance (3.9 nm) from the Pb islands at three types of sites, A, B, and C. The averaged values for each site are $-0.38 \pm 0.08$ (A, red), $-0.60 \pm 0.05$ (B, blue) and $-0.89 \pm 0.06$ (C, green). The number of sampling points are 229, 73 and 54 for A, B and C, respectively.

**S4. Density of states in the confined area**

In order to eliminate the possibility that the enhanced proximity effect observed in narrow terraces of the SIC phase is caused by the modulation of the DOS in the SIC phase, we measured the spatial distribution of ZBC under the magnetic field of 0.17 T. The applied magnetic field is strong enough to weaken the superconductivity in the periphery of the Pb islands and to suppress the proximity effect. The measured ZBC, therefore, corresponds to the local density of state (LDOS) of the SIC phase. Since there is no modulation in DOS or standing waves observed in the image, the enhanced proximity effect is not due to the DOS modulation.

On the Pb islands, vortices are observed due to the magnetic field penetrations through the islands

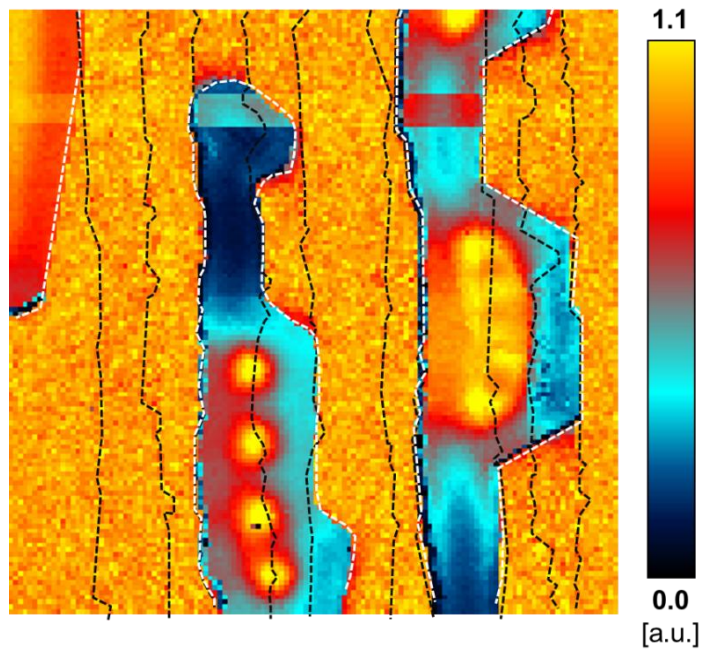

**Figure S4.** 1.0 μm × 1.0 μm ZBC map of the same area as figure 1(a) under a magnetic field of 0.17 T applied perpendicular to the substrate. The circular objects in the figure are vortices.

## S5. The Usadel equation and numerical results

The Usadel equation has been used before to study the proximity effect in a planar superconductor/normal metal interface and yields a satisfactory description. Here we consider one-dimensional case and the Usadel equation using a complex parameter $\theta_{S(N)}$ can be written as

$$\frac{\hbar D_{S(N)}}{2} \partial_x^2 \theta_{S(N)} + iE \sin\theta_{S(N)} + \Delta \cos\theta_{S(N)} = 0, \quad [12]$$

where $E$ is the energy relative to the Fermi energy, $D_{S(N)}$ is the diffusion constant of the superconductor (normal metal), and $\Delta$ is the energy gap, which is zero in normal metals. The boundary condition at the superconductor / normal interface is

$$\sigma_N \partial \theta_N = \sigma_S \partial \theta_S = g_{SN}(\theta_S - \theta_N),$$

where $\sigma_{S(N)}$ is the conductivity of the superconductor just above $T_c$ (normal metal) and $g_{SN}$ is the interface conductivity, which characterizes the electronic transparency of the interface. In a narrow SIC phase region, the superconducting correlation is nonzero at the other boundary of the SIC phase, and the corresponding boundary condition becomes important. We introduce the superconductor/normal metal/normal metal (SNN) model, as shown in Fig. 3(c), to describe the diffusion of electrons in the narrow SIC region. The interfacial conductivity between two normal metals $g_{NN}$ parametrizes the energy barrier at the interface between the two normal metals. The corresponding boundary at the interface is

$$\sigma_N \partial \theta_{N1} = \sigma_N \partial \theta_{N2} = g_{NN}(\theta_{N2} - \theta_{N1}).$$

We calculate the LDOS= $N_0 \operatorname{Re}\left[\cos\theta_{S(N)}(x,E)\right]$ at the Fermi energy $E = 0$ and fit it to the ZBC measured experimentally, from which we obtain $D_{S(N)}$, $\sigma_{S(N)}$, $g_{SN}$, and $g_{NN}$.

By fitting to the measured ZBC in the wide SIC region as shown in Fig. S2, where superconducting correlation decays to zero, we obtained $D_S = 30$ cm$^2$/s by using $\Delta = 1.08$ meV. The fitted $D_S$ is in reasonable agreement with the value $D_S = v_{F,S} l_S / 3 \sim 47$ cm$^2$/s with the Fermi velocity $v_{F,S} \sim 10^6$ m/s and the mean free path $l_S = 3.5t \sim 14$ nm for the 14-ML Pb island with thickness $t = 4.0$ nm. The fitted diffusion constant in the SIC phase is $D_N = 1.2$ cm$^2$/s. By using the relation $D_N = v_F \cdot l/2$ and $v_F = \hbar k_F / m_e^*$, where $v_F$ is the Fermi velocity, $l$ is the mean free path, $k_F$ is the Fermi wave number and $m_e^*$ is the effective electron mass of the metallic SIC state, and experimentally obtained $k_F = 13.6$ nm$^{-1}$ and $m_e^* = 1.16\, m_e$,[8] the mean free path in the SIC phase is estimated at 1.8 nm. The relation $l_N \ll \xi$, where $\xi$ is the coherence length of the normal metal given by $\sqrt{D_N \hbar / 2\Delta}$, justified the dirty limit, where the Usadel equation is valid. The conductivity of the Pb island just above $T_c$ is $\sigma_S = 10^7$ $(\Omega \cdot m)^{-1}$. According to the fitting, we estimate the conductivity in the SIC phase as $\sigma_N = 2 \times 10^6$ $(\Omega \cdot m)^{-1}$.

We then fitted the measured ZBC in the narrow SIC regions by using $D_S$, $D_N$, $\sigma_S$ and $\sigma_N$ obtaind in the aforementioned fitting. The fitting parameters are $g_{SN}$ and $g_{NN}$ in this case. The fitting for an opaque interface at two metals with $g_{NN} = 0$, which corresponds to the case of SN model in Fig. S5(b), are shown by dotted lines in Fig. S5(a). However this fitting overestimate the induced superconducting correlation in the SIC region, which implies that the interface at the two normal metals are transparent. The transmission of the electrons in surface states through the step edges has also been observed before [13]. We then performed the fitting by using $g_{NN}$ as another parameter and the experimental data can be fitted satisfactorily as shown in Fig. S5(a). The fitted $g_{NN}$ and $g_{SN}$ are compiled in table I. The nonzero transmission also induces the proximity effect across the step edge. However the calculated ZBC over the step edge, which is shown in Fig. S5(c), is quite small because of the small $g_{NN}$, which is also consistent with the observed termination of the proximity effect by the step edges.

Table I

| terrace width (L) [nm] | $g_{SN} \times 7 \times 10^{13} (\Omega \cdot m)^{-1}$ | $g_{NN} \times 7 \times 10^{13} (\Omega \cdot m)^{-1}$ |
|---|---|---|
| 76.7 | 0.55 | 0.10 |
| 54.3 | 0.56 | 0.11 |
| 38.3 | 0.56 | 0.19 |
| 25.5 | 0.80 | 0.26 |
| 12.8 | 0.64 | 0.20 |

The averaged sheet conductivity at the step edge of the two normal metals is $g_{NN,2D} = g_{NN} \cdot t_{SIC} \sim 4000$ $(\Omega \cdot m)^{-1}$, which is consistent with the value estimated by Uchihashi *et al.* via the Josephson current measurements [14]. Here $t_{SIC}$ is the thickness of the SIC phase, which is assumed to be 1 bilayer of the Si(111) substrate (0.30 nm).

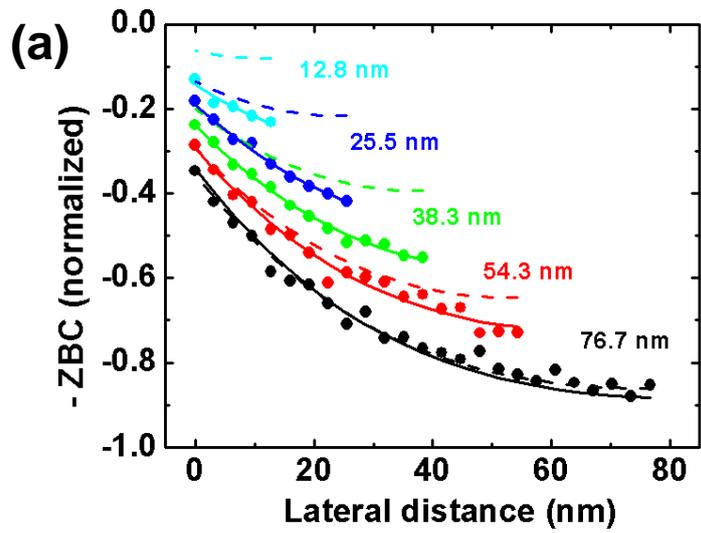
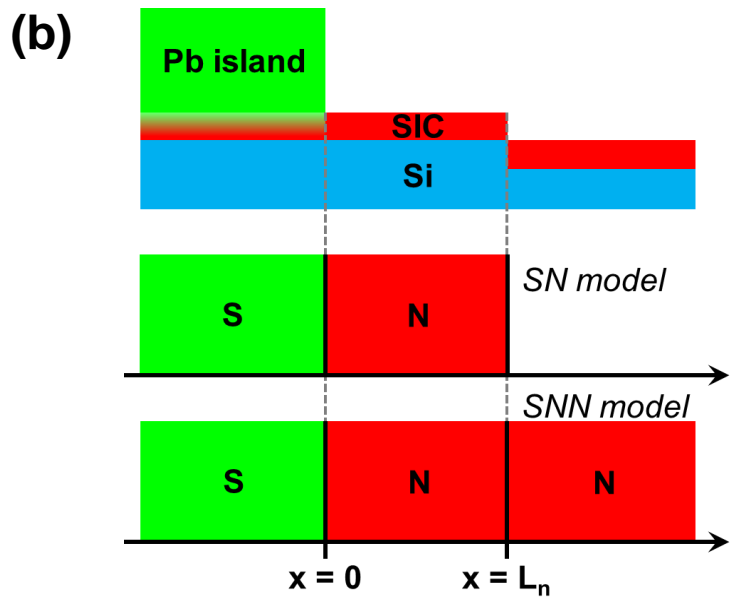
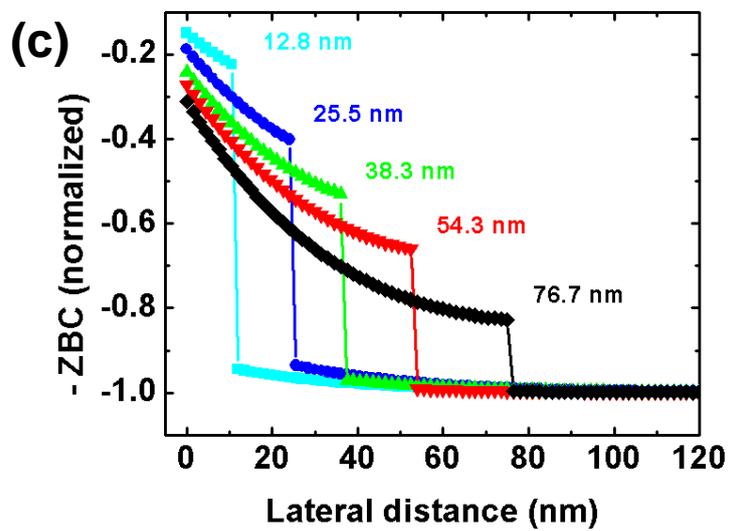

**Figure S5.** (a) Fitting results of the ZBC profiles based on the Usadel equations for SN (dotted lines) and SNN models (solid lines). (b) Schematics of SN model and SNN model (c) normalized negative ZBC profiles across the NN junctions with different length of the normal metals calculated with the Usadel equation for the SNN model.